# STAN: SMALL TUMOR-AWARE NETWORK FOR BREAST ULTRASOUND IMAGE SEGMENTATION

*Bryar Shareef, Min Xian\*, Aleksandar Vakanski*

Department of Computer Science, University of Idaho

## ABSTRACT

Breast tumor segmentation provides accurate tumor boundary, and serves as a key step toward further cancer quantification. Although deep learning-based approaches have been proposed and achieved promising results, existing approaches have difficulty in detecting small breast tumors. The capacity to detecting small tumors is particularly important in finding early stage cancers using computer-aided diagnosis (CAD) systems. In this paper, we propose a novel deep learning architecture called Small Tumor-Aware Network (STAN), to improve the performance of segmenting tumors with different size. The new architecture integrates both rich context information and high-resolution image features. We validate the proposed approach using seven quantitative metrics on two public breast ultrasound datasets. The proposed approach outperformed the state-of-the-art approaches in segmenting small breast tumors.

*Index Term*— breast ultrasound, small tumor segmentation, deep learning, multi-scale features, STAN

## 1. INTRODUCTION

According to the National Center for Health Statistics [1], in 2019, United States is expected to have 891,480 new women cancer cases, where 30% of the all cases will be breast cancer. Early detection is the key to improving the survival rate of breast cancer; the five-year relative survival rate is 98% if the breast cancer is detected and treated at the early stages, and only 22% in cases with advanced-stage cancers. Computer-aided diagnosis (CAD) systems have been proposed to detect breast cancer automatically. In these systems, breast tumor segmentation is a key step that help accurate tumor quantification. Tremendous number of breast tumor segmentation approaches have been proposed in the last two decades; and some approaches have achieved promising overall performance on their private datasets. However, most approaches cannot segment small tumors accurately. Breast ultrasound (BUS) images are used in this study since ultrasound imaging is noninvasive, painless, nonradioactive and cost-effective.

\* Correspondence to Min Xian (mxian@uidaho.edu). This work was supported, in part, by the Center for Modeling Complex Interactions (CMCI) at the University of Idaho through NIH Award #P20GM104420.

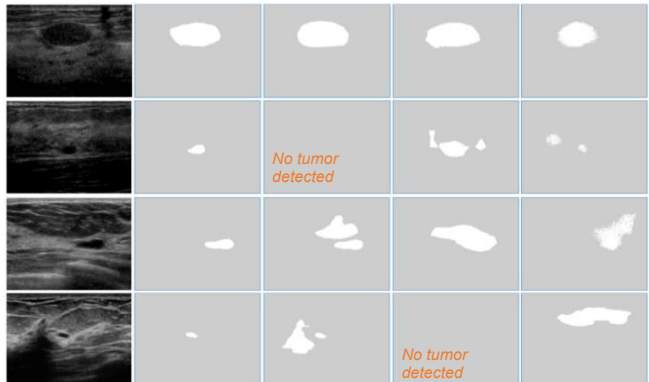

(a) BUS images (b) Ground truth (c) U-Net [15] (d) AlexNet [19] (e) SegNet [18]

**Fig. 1.** Performance of state-of-the-art approaches for segmenting breast tumors with different sizes.

In the last two decades, breast tumor segmentation has been an active research area. Existing approaches can be classified into traditional approaches and deep learning approaches. Various traditional image processing approaches have been applied to BUS image segmentation, such as thresholding [2-5], region growing [6,7], and watershed [8]. However, the traditional methods are not robust due to poor scalability and sensitivity to noise. Refer to [20] for a detailed review of BUS segmentation approaches.

Deep learning approaches [9-12,21] have recently demonstrated state-of-the-art performance for breast ultrasound segmentation. Cheng *et al.* [10] employed a stacked denoising auto-encoder (SDAE) to diagnose breast ultrasound lesions and lung CT nodules. The information extension strategy was used in [11], where the wavelet feature was added to the original image to train a fully convolutional network (FCN). Breast anatomy information was applied to the Conditional Random Fields (CRFs) to enhance the segmentation performance. In addition, Huyanh *et al.* [12] used transfer learning for classification of BUS images, however, the proposed model does not perform tumor segmentation. Similarly, Yap *et al.* [9] used three different deep learning methods, a patch-based LeNet, a U-Net, and a transfer learning approach with a pre-trained FCN-AlexNet on two different datasets to segment BUS images. However, they failed to achieved good performance for segmenting small tumors. Furthermore, a very deep CNN architecture GoogleNet In-ception v2 in [13] is used for the classification task, to distinguish between benign and malignant tumors. The results showed that the CNN model had better or equal diagnostic performance compared to radiologists. Moreover, in order to focus on regions with high saliency values, the method in [21] integrates radiologists' visual attention for BUS segmentation.

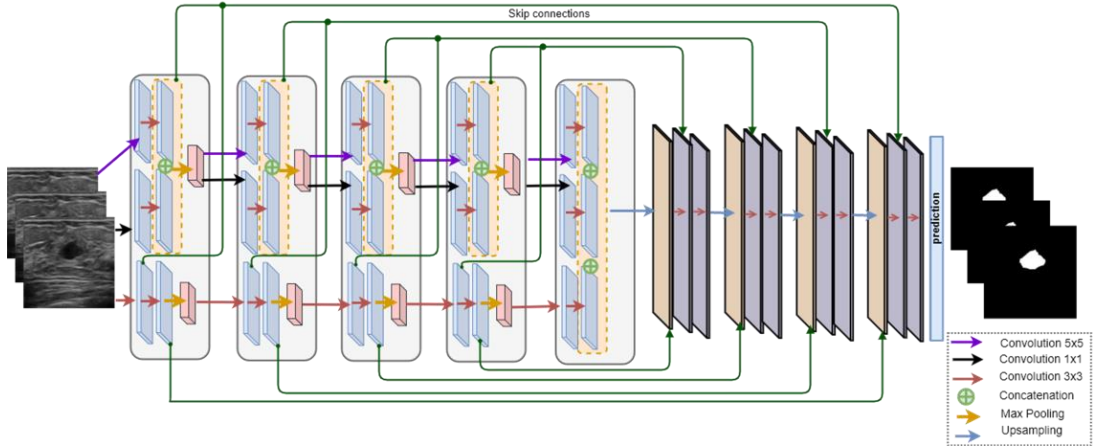

**Fig. 2.** The STAN architecture. The block sizes do not represent the actual feature maps.

In this paper, our results indicate that the three state-of-art models (FCN-AlexNet, SegNet, and regular Unet) have difficulty in detecting small tumors (Fig.1). We propose a novel architecture based on the core of U-Net architecture to solve the current issue of segmenting small tumors in breast ultrasound images. The method is validated using two public datasets. The experimental results demonstrate enhanced ability of the proposed model for small tumor detection in comparison to existing methods.

## 2. PROPOSED METHOD

The proposed method is based on one key observation: the size of breast tumors varies dramatically among patients; and existing deep neural networks that use fixed kernel size cannot detect small breast tumors accurately. To overcome this problem, we propose the Small Tumor-Aware Network (STAN) to extract and fuse image context information at different scales. STAN constructs feature maps using kernels with three different sizes at each convolutional layer in the encoder. Such feature maps carry multiscale context information and preserve fine-grained tumor location information. Consequently, STAN improves the performance of breast tumor segmentation, especially for small tumors. Fig. 2 illustrates the overall architecture of STAN.

### 2.1 STAN Architecture

The size of the receptive field is a crucial issue in deep neural networks, because the output must response to an appropriate size of regions to capture objects with different sizes. There are two main ways to tune the size of the receptive field: 1) downsampling; and 2) stacking more layers. The two methods can only increase the receptive field, and are suitable for segmenting large objects. In BUS image segmentation, a large receptive field will result in high false positives. Therefore, our goal is to avoid stacking too many layers with large kernel size, and design an architecture that has different sizes of the receptive field.

The proposed approach has a similar architecture as the general U-Net: i.e., it contains a contracting and expanding stage with skipping links. Unlike the U-Net architecture, where the contracting stage has only one branch, the proposed network comprises two encoder branches. In addition, the proposed network has three skipping links (the green links in Fig. 2) between the encoder and decoder blocks, which allows retaining and propagating high-resolution features to the decoder. E.g., for the $i$th block, we denote the output of the two encoder branches as $C_{i,1}$ and $C_{i,2}$, and the next block will output

$$C_{i+1,1} = p\left(conv_3\left(conv_3(C_{i,1})\right)\right) \quad (1)$$

$$C_{i+1,2} = p\left(conv_3\left(conv_1(C_{i,2})\right) \oplus conv_3\left(conv_5(C_{i,2})\right)\right) \quad (2)$$

where $conv_n$ denotes the convolutional operation with kernel size $n \times n$. $C_{0,1}$ and $C_{0,2}$ are used to denote an input image to the network, where $C_{0,1} = C_{0,2}$; $p$ denotes the max pooling operation; and, for the central layer, $C_{5,1}$ and $C_{5,2}$ are

$$\begin{aligned}C_5 = C_{5,1} = C_{5,2} = &\ conv_5\left(conv_5(C_{4,1})\right) \oplus \\ &\ conv_1\left(conv_1(C_{4,2})\right) \oplus \\ &\ conv_3\left(conv_3(C_{4,2})\right).\end{aligned} \quad (3)$$

In Eqs. (1-3), $\oplus$ denotes the concatenation operation. From the blocks one to four, each block applies kernels with three different sizes, that is 1×1, 3×3 and 5×5, and captures image features at three different scales. In general, when the dimensions of the input images to the neural network are reduced extremely via down-sampling layers, the network performs poorly because the network loses vast amount of information, recognized as a representational bottleneck [14]. To solve the representational bottleneck issue, the network-in-network architecture [14] used convolutional kernels of size 1×1 followed by a ReLU layer to introduce more no-linearity. Motivated by this approach, in the second branch of the encoder, we introduced 1×1 kernels to increase the representational power of the model.

Table 1. Segmentation performance of four approaches on two datasets.

| Datasets | Methods | TPR | FPR | JI | DSC | AER | AHE | AME |
|---|---|---|---|---|---|---|---|---|
| BUSIS | FCN AlexNet | **0.950** | 0.336 | 0.736 | 0.841 | 0.386 | 25.1 | 7.1 |
| | SegNet | 0.938 | 0.158 | 0.820 | 0.895 | 0.220 | 21.7 | 4.5 |
| | U-Net | 0.920 | 0.138 | 0.825 | 0.897 | 0.218 | 26.8 | 4.9 |
| | STAN | 0.917 | **0.093** | **0.847** | **0.912** | **0.176** | **18.9** | **3.9** |
| Dataset B | FCN AlexNet | **0.868** | 1.167 | 0.469 | 0.610 | 1.299 | 40.8 | 14.5 |
| | SegNet | 0.852 | 0.834 | 0.595 | 0.708 | 0.982 | 41.6 | 11.4 |
| | U-Net | 0.776 | 0.406 | 0.653 | 0.745 | 0.630 | 39.6 | 10.8 |
| | STAN | 0.801 | **0.266** | **0.695** | **0.782** | **0.465** | **35.5** | **9.7** |

The original U-Net architecture copies features after the second convolutional layer in the encoder part and concatenates the features to the corresponding layer in the decoder section. In our proposed model, the skipping links involve the output of the first convolution in each layer merged to the result of the first convolution in the corresponding decoder part. In addition, a skipping layer from the merging of the two new layers after the second convolution in the encoder merges to the result of the second convolution in the decoder part. Accordingly, the expanding stage is enriched by fusing feature maps from the blocks in the two encoders. Let $U_i$ ($i = 5, 4, 3, 2, 1$) be the output of $i$th up-sampling block; and the output of the next bock is given by

$$U_{i-1} = conv\left(conv\left(DeConv(U_i \oplus C_{i-1,1})\right)\right) \oplus conv_5(C_{i,1}) \oplus C_{i-1,2} \quad (4)$$

In Eq.(4), $U_5$ is equal to $C_5$ from the central layer, and $DeConv$ denotes the deconvolution operation. In addition, since the layer five does not involve pooling, we discarded the pooling layers from the skipping block. The original skipping layers stay the same, where we combine it to the up-sampling layer before the first convolutional layer.

**2.2 Implementation and Training**

The input images and their corresponding ground truths are resized to 256×256. Since the datasets are of small size, we applied image width and height shift to augment the training set. The batch size is 4, and the number of training epochs is set to 50. Adam optimizer [9] is utilized for training the proposed network, and the initial learning rate is set to 0.0001.

Let $P = \{p_i\}_{i=1}^{N}$ and $G = \{g_i\}_{i=1}^{N}$ be the output of the final pixel-wise sigmoid layer and the ground truth, respectively. The loss function is computed by using discrete dice loss [16]:

$$L_{dice} = 1 - \frac{1 + 2\sum_i^N p_i g_i}{1 + \sum_i^N p_i^2 + \sum_i^N g_i^2}$$

## 3. EXPERIMENTAL RESULTS

### 3.1. Dataset, metrics and setup

We use two publicly available datasets to validate the performance of the proposed approach, BUSIS dataset [17] and Dataset B [9]. The BUSIS dataset contains 562 images from three hospitals using GE VIVID 7, LOGIQ E9, Hitachi EUB-6500, Philips iU22, and Siemens ACUSON S2000. The Dataset B has 163 breast ultrasound images, and the UDIAT Diagnostic Centre of the Parc Taulı́ Corporation, Sabadell (Spain) collected the images using Siemens ACUSON Sequoia C512 system with 17L5 linear array transducer.

Both area and boundary metrics are used to evaluate the segmentation results. The metrics are true positive ratio (TPR), false positive ratio (FPR), Jaccard index (JI), dice's coefficient (DSC), area error ratio (AER), Hausdorf error (HE) and mean absolute error (MAE). The performance of the proposed method is compared with the SegNet [18], FCN-AlexNet [19], and U-Net [15]. The FCN-AlexNet is pre-trained using the ImageNet, and all other approaches are trained from scratch. We employ 5-fold cross-validation to evaluate the test performance of all methods.

### 3.2. Overall Performance

The overall quantitative results are shown in Table 1, where the proposed STAN method outperformed the other three approaches in six metrics on the two datasets. FCN-AlexNet, SegNet, and U-Net produced high TPRs on the BUSIS dataset, and FCN-AlexNet and SegNet obtained higher TPRs than the proposed approach on the Dataset B. However, they achieved high TPR at the cost of large false positive ratio (FPR) shown in the fourth column of Table 1.

Fig. 3 compares the segmentation results of SegNet, FCN-AlexNet, U-Net, and the proposed STAN. Fig. 3(b)

Table 2. Small Tumor Segmentation.

| Dataset | Method | TPR | FPR | JI | DSC | AER | AHE | AME |
|---|---|---|---|---|---|---|---|---|
| BUSIS | FCN-AlexNet | **0.947** | 0.767 | 0.603 | 0.732 | 0.821 | 26.3 | 9.6 |
|  | SegNet | 0.923 | 0.251 | 0.747 | 0.841 | 0.328 | 22.4 | 6.2 |
|  | U-Net | 0.920 | 0.296 | 0.756 | 0.843 | 0.376 | 44.2 | 8.3 |
|  | STAN | 0.902 | **0.165** | **0.791** | **0.870** | **0.263** | **21.3** | **5.2** |
| Dataset B | FCN-AlexNet | **0.868** | 1.863 | 0.353 | 0.492 | 1.995 | 49.2 | 18.4 |
|  | SegNet | 0.854 | 1.452 | 0.495 | 0.619 | 1.598 | 50.1 | 14.2 |
|  | U-Net | 0.768 | 0.682 | 0.593 | 0.681 | 0.913 | 43.1 | 13.8 |
|  | STAN | 0.814 | **0.400** | **0.673** | **0.759** | **0.586** | **35.9** | **11.1** |

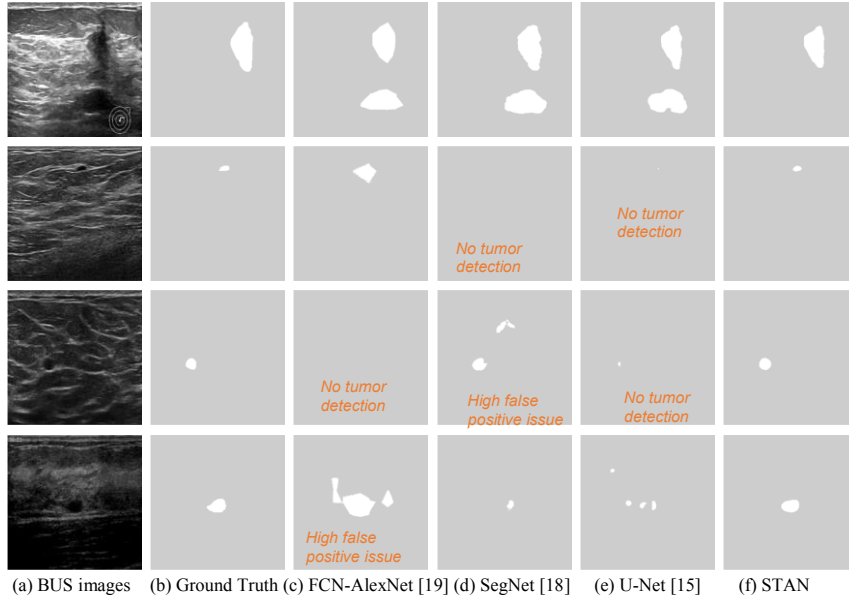

(a) BUS images  (b) Ground Truth  (c) FCN-AlexNet [19]  (d) SegNet [18]  (e) U-Net [15]  (f) STAN

**Fig. 3.** Small tumor segmentation.

shows the corresponding ground truth of the original BUS images in Fig. 3(a). As shown in the first row, FCN-AlexNet, SegNet, and U-Net produce high false positives, while the proposed STAN can accurately segment the tumors. In the second row of Fig. 3, the FCN-AlexNet has high false positives compared to the ground truth; and both the SegNet and U-Net fail to detect the tumor.

### 3.3. Small Tumor Segmentation

In this section, we evaluate the performance of four approaches in segmenting small tumors. The criterion to select small tumors is the length of the longest axis of a tumor region, and the length threshold is set to 120 pixels. The physic sizes of tumors are not used because they are unavailable for most images in the two datasets. 76 and 49 images are selected form the BUSIS and Dataset B, respectively.

As shown in Table 2, on the two datasets, all metrics except the TPR of the proposed STAN are better than those of FCN-AlexNet, SegNet, and U-Net. The FPR of the FCN-AlexNet on the small dataset (**0.767**) of is more than twice as its original FPR in Table 1(**0.336**). All other three approach generate high FPRs (FCN-AlexNet: 1.86, SegNet: 1.45 and U-Net: 0.68) for small tumors in the Dataset B. The third and fourth rows of Fig. 3 show segmentation results of a small tumor, the FCN-AlexNet and U-Net detect no tumor; while the SegNet produced high false positive. In the fourth row, the FCN-AlexNet and U-Net generated high false positive, and the SegNet only found a small part of the tumor.

### 4. CONCLUSION

In this paper, we proposed the Small Tumor-Aware Network (STAN) to overcome challenges in breast tumor early detection. The STAN has two encoder branches that extract and fuse image context information at different scales. The model constructs feature maps using kernels with three different sizes at each convolutional layer. These feature maps carry multiscale context information and preserve fine-grained tumor location information. The proposed STAN achieved the state-of-the-art overall performance on two public datasets, and outperformed the other three segmentation approaches in segmenting small tumors.

In the future, we will focus on improving the robustness of the proposed STAN.